\documentclass[aps,prd,nofootinbib]{revtex4}
\usepackage{amsmath,amssymb}
\usepackage[hyperfootnotes=false]{hyperref}

\begin{document}

\title{Ultralight dark photon and Casimir effect}

\author{Abdaljalel Alizzi}
\email{abdaljalel90@gmail.com}\affiliation{Novosibirsk State University, 
Novosibirsk 630 090, Russia}

\author{Z.~K.~Silagadze}
\email{Z.K.Silagadze@inp.nsk.su}\affiliation{Budker Institute of
Nuclear Physics and Novosibirsk State University, Novosibirsk 630
090, Russia }

\begin{abstract}
We investigate the influence of a dark photon on the Casimir effect and
calculate the corresponding leading contribution to the Casimir energy. For 
expected magnitudes of the photon - dark photon mixing parameter, the 
influence turns out to be negligible. The plasmon dispersion relation is
also not noticeably modified by the presence of a dark photon.
\end{abstract}


\maketitle

\section{introduction}
The true nature and composition of the dark matter is currently unknown,
which remains one of the most significant unsolved problems in modern physics. 
Dark photons (also known as hidden, para-, or secluded photons) are 
well-motivated dark matter candidates \cite{1}. Dark photon is the gauge boson
of the additional $U(1)$ gauge group beyond the Standard Model. What makes
such extra $U(1)$ gauge factors of particular phenomenological interest is 
that they are a generic feature of string theory \cite{2}.

Dark photons can kinetically mix with the Standard Model photon \cite{3}.
This makes it possible to search for them in astrophysical and cosmological 
observations and in laboratory experiments \cite{1,4}. Below the electroweak 
scale, the minimum coupling between a dark photon and particles of the 
visible sector (Standard Model particles) can be described by the following 
Lagrangian density 
\begin{equation}
{\cal{L}}=-\frac{1}{4}F_{a\mu\nu}F^{a\mu\nu}-\frac{1}{4}F_{b\mu\nu} 
F^{b\mu\nu}+\frac{1}{2}m_b^2A_{b\mu} A^{b\mu}-
\frac{\epsilon}{2}
F_{a\mu\nu}F^{b\mu\nu}-e_aJ_\mu A^{a\mu}-e_b \tilde J_\mu A^{b \mu},
\label{eq1}
\end{equation}
where the gauge boson $A_a^\mu$ couples to the ordinary electromagnetic current
$J_\mu$, $A_b^\mu$ couples to the dark sector analog of the electromagnetic 
current $\tilde J_\mu$ (if any), and $\epsilon$ is the dimensionless kinetic 
mixing parameter, which in some scenarios can be as large as $10^{-3}-10^{-2}$
\cite{5}. The dark boson mass, $m_b$, can arise, for example by 
St\"{u}ckelberg mechanism \cite{6}.

The Lagrangian (\ref{eq1}) can be diagonalized by introducing the physical 
fields (mass eigenstates) $A_\mu$ and $\tilde A_\mu$:
\begin{equation}
A_{a\mu}=A_\mu-\frac{\epsilon}{\sqrt{1-\epsilon^2}}\,\tilde A_\mu,\;\;\;
A_{b\mu}=\frac{1}{\sqrt{1-\epsilon^2}}\,\tilde A_\mu,
\label{eq2}
\end{equation}
in terms of which the Lagrangian (\ref{eq1}) takes the form
\begin{equation}
{\cal{L}}=-\frac{1}{4}F_{\mu\nu}F^{\mu\nu}-\frac{1}{4}\tilde F_{\mu\nu}\tilde 
F^{\mu\nu}+\frac{1}{2}\,\frac{m_b^2}{1-\epsilon^2}\,\tilde A_{\mu} \tilde A^{\mu}-
e_aJ_\mu A^{\mu}-\frac{1}{\sqrt{1-\epsilon^2}}\left (e_b \tilde J_\mu -
\epsilon e_a J_\mu\right )\tilde A^{\mu}.
\label{eq3}
\end{equation}
As we see, the dark photon $\tilde A_\mu$ can interact with ordinary charged
particles. If the dark photon mass $\mu=\frac{m_b}{\sqrt{1-\epsilon^2}}$ is
very small (as we assume), then the measured Coulomb force between two 
charged particles will be $F_c=\frac{e_a^2}{4\pi r^2}\left (1+
\frac{\epsilon^2}{1-\epsilon^2}\right )=\frac{e_a^2}{4\pi r^2}\,\frac{1}
{1-\epsilon^2}$. Therefore the measured electric charge is 
$e=\frac{e_a}{\sqrt{1-\epsilon^2}}$, and (\ref{eq3}) expressed in terms of the 
renormalized  quantities (the charge $e$ and the mass $\mu$) reads
\begin{equation}
{\cal{L}}=-\frac{1}{4}F_{\mu\nu}F^{\mu\nu}-\frac{1}{4}\tilde F_{\mu\nu}\tilde 
F^{\mu\nu}+\frac{1}{2}\mu^2\tilde A_{\mu} \tilde A^{\mu}- 
eJ_\mu \left ( \sqrt{1-\epsilon^2} A^{\mu}-\epsilon \tilde A^{\mu} \right )
-\frac{e_b}{\sqrt{1-\epsilon^2}}\tilde J_\mu\tilde A^{\mu}.
\label{eq4}
\end{equation}
Such kind of phenomenological model (without $\tilde J_\mu$, and without any
connection with the vector bosons mixing) was first investigated by Okun 
\cite{7}, and consequences of the mixing of two abelian gauge bosons were 
first illustrated by Holdom \cite{3}.

The combination $A^\prime_\mu=\sqrt{1-\epsilon^2} A^{\mu}-\epsilon \tilde A^{\mu}$,
which has a direct coupling to the electromagnetic current, acts as an active 
photon. The orthogonal combination $\tilde A^\prime_\mu=\epsilon A^{\mu}+
\sqrt{1-\epsilon^2}\tilde A^{\mu}$ acts as a sterile photon, which has 
no direct coupling to the electromagnetic current. The transformation 
\begin{equation}
\left ( \begin{array}{c} A_\mu \\ \tilde A_\mu \end{array} \right )=
\left ( \begin{array}{cc}\sqrt{1-\epsilon^2} & \epsilon \\
-\epsilon & \sqrt{1-\epsilon^2} \end{array} \right )
\left ( \begin{array}{c} A^\prime_\mu \\ \tilde A^\prime_\mu \end{array} \right )
=\left ( \begin{array}{rr}\cos{\theta} & \sin{\theta} \\
-\sin{\theta} & \cos{\theta} \end{array} \right )
\left ( \begin{array}{c} A^\prime_\mu \\ \tilde A^\prime_\mu \end{array} \right )
\label{eq5}
\end{equation}
is a rotation which leaves the kinetic terms in (\ref{eq4}) diagonal, but 
induces a mass mixing between the rotated fields:
\begin{equation}
{\cal{L}}=-\frac{1}{4}F^\prime_{\mu\nu}F^{\prime\mu\nu}-\frac{1}{4}\tilde 
F^\prime_{\mu\nu}\tilde F^{\prime\mu\nu}+ \frac{1}{2}\mu^2\cos^2{\theta}\left (
\tilde A^{\prime}_\mu-\tan{\theta}A^{\prime}_\mu\right ) 
\left (\tilde A^{\prime\mu}-\tan{\theta}A^{\prime\mu}\right )- 
eJ_\mu A^{\prime\mu}-e_b\tilde J_\mu \left (\tilde A^{\prime\mu}-\tan{\theta}
A^{\prime\mu}\right ).
\label{eq6}
\end{equation}
In terms of  $A^\prime_\mu$ and $\tilde A^\prime_\mu$, the original $A_{a\mu}$
and $A_{b\mu}$ fields are expressed as follows
\begin{equation}
A_{a\mu}=\frac{1}{\sqrt{1-\epsilon^2}}A^\prime_\mu,\;\;\;
A_{b\mu}=\tilde A^\prime_\mu-\frac{\epsilon}{\sqrt{1-\epsilon^2}}A^\prime_\mu.
\label{eq7}
\end{equation}
In analogy with neutrino oscillations, $A^\prime_\mu$ and $\tilde A^\prime_\mu$
fields can be called flavor eigenstates, while  $A_\mu$ and $\tilde A_\mu$ are
mass eigenstates (propagating states). 

Naively, one can expect that the resulting  physics does not depend on the 
choice of the basis  (\ref{eq2}) or (\ref{eq7}), and such a choice is a matter 
of convenience \cite{9,9A}. This is indeed the case in quantum mechanics. 
However, in quantum field theory, the situation is more subtle, since the 
Fock spaces for fields with definite mass and fields with definite flavor are 
not unitarily equivalent \cite{10,11,12}.

In light of this inequivalence between mass and flavor vacua, it is interesting
to ask how dark photons and the corresponding mixing phenomena affect the 
Casimir energy. At first glance, a natural choice for studying the effect of
field mixing on the Casimir energy is the propagating basis (\ref{eq2}) and the 
corresponding Lagrangian (\ref{eq4}). In this note we make such a choice and 
try to investigate the influence of photon - dark photon mixing on the Casimir 
energy. Natural units ($\hbar=1,\,c=1$) are assumed, unless $\hbar$ and/or $c$
are explicitly indicated in the equation. For electromagnetic quantities we
use Lorentz-Heaviside units.

\section{Casimir effect}
Casimir effect is a fascinating phenomenon (Pauli initially dismissed it as 
``absolute nonsense'' \cite{13A}) in which two electrically neutral, 
parallel metal plates, spaced a short distance from each other, experience 
a force of mutual attraction. After its first discovery by Hendrik Casimir
\cite{13},  there is still debate about whether the Casimir effect is 
a manifestation of the reality of zero-point quantum fluctuations of the 
electromagnetic vacuum or is it just the  relativistic, retarded van der Waals 
force between the metal plates \cite{14}. Casimir himself, when asked the same 
question twice with an interval of eleven years, whether the Casimir effect is 
the result quantum fluctuations of the electromagnetic field, or is it caused 
by van der Waals forces between molecules in two objects, replied that he
has not made up his mind (see foreword by I.H. Brevik in \cite{15}).

Nevertheless, since pioneering works by Casimir and Polder, the normal-mode 
expansion of a quantized electromagnetic field inside a cavity with suitable 
boundary conditions has been widely used to study dispersion interactions 
\cite{16} to which the Casimir force belongs. The picture of a fluctuating 
quantum vacuum, altered by the presence of an object, has proven to be very 
useful for predicting new and interesting effects such as the quantum 
atmosphere effect \cite{17} and the dissipationless rotation-induced axial 
Casimir force that emerges when a particle rotates above a plate that 
exhibits either time-reversal symmetry breaking or parity-symmetry breaking 
\cite{18}. Therefore, we consider it appropriate to remind the spirit of the 
Casimir's original derivation \cite{13}. 

To quantize the electromagnetic field, a large cubic box of size L with 
periodic boundary conditions is usually considered. Then the wave vector
$\vec{k}$ of the electromagnetic field is quantized inside the box:
\begin{equation}
k_x=\frac{2\pi}{L}\,n,\;\;\;k_y=\frac{2\pi}{L}\,m,\;\;\;k_z=\frac{2\pi}{L}\,l,
\;\;\; n,m,l=0,\pm 1, \pm 2,\ldots
\label{eq8}
\end{equation}
Zero-point energy of the fluctuating electromagnetic vacuum inside the box is
\begin{equation}
W_L=\frac{1}{2}\sum\limits_{\vec{k},\,\sigma}\hbar\omega_k=\hbar c
\sum\limits_{\vec{k}}\sqrt{k_x^2+k_y^2+k_z^2}\;{\underset{L\to\infty}
\longrightarrow}\;\hbar c\,\frac{L^3}
{(2\pi)^3}\int d\vec{k}\,\sqrt{k_x^2+k_y^2+k_z^2},
\label{eq9}
\end{equation}
where $\sigma$ labels two photon polarizations, and for
$L\to\infty$ we can replace the sum by the integral according to the rule
$\frac{1}{L^3}\sum\limits_{\vec{k}}\to\int\frac{d\vec{k}}{(2\pi)^3}$ since the
number of quantized $\vec{k}$-states per $d\vec{k}$ according to (\ref{eq8}) 
is $$\frac{dN}{d\vec{k}}=\frac{dn\,dm\,dl}{dk_x\,dk_y\,dk_z}=
\frac{L^3}{(2\pi)^3}.$$

If two ideally conducting $L\times L$ plates are inserted at $x=0$ and $x=R$ 
in the previous construction, the vacuum energy will change, since the plates 
impose Dirichlet boundary conditions at their locations, and, correspondingly, 
the quantization conditions of the zero-point electromagnetic field between 
the plates will change to
\begin{equation}
k_x=\frac{\pi}{L}\,n,\;\;\;k_y=\frac{2\pi}{L}\,m,\;\;\;k_z=\frac{2\pi}{L}\,l,
\;\;\; n=0,1,2,\ldots, \;\; m,l=0,\pm 1, \pm 2,\ldots
\label{eq10}
\end{equation}
As before, we can replace the summations over $k_y$ and $k_z$ with the 
corresponding integrals in the limit $L\to\infty$. However, the $k_x$ sum 
should be kept as the $R$ distance between the plates remains finite. 

There is one subtlety when considering this sum (see, for example, \cite{19}). 
As already mentioned, in the general case we have two independent 
polarizations of an electromagnetic wave. The corresponding modes are called 
transverse electric (TE) and transverse magnetic (TM), respectively. In TE
mode, the electric field is parallel to the plates, and in TM mode, the 
magnetic field is parallel to the plates. There is no TE mode that corresponds 
to $n=0$, since in this case the boundary conditions will force the electric 
field  between the plates to be identically zero. 

Therefore, for the zero-point energy of the fluctuating electromagnetic field
between the conducting plates, we have  
\begin{equation}
W_R=\frac{\hbar c L^2}{(2\pi)^2}\sum\limits_{n=(0)1}^\infty\iint dk_y\,dk_z
\sqrt{k_y^2+k_z^2+\left(\frac{\pi n}{R}\right)^2},
\label{eq11}
\end{equation} 
where the notation $\sum\limits_{n=(0)1}^\infty$ means that the term $n=0$ in the
sum must be multiplied by $1/2$ \cite{13}. 

When calculating the zero-point energy of the fluctuating electromagnetic field
outside the conducting plates (but inside the quantization volume $L^3$) 
$W_{L-R}$, we should specify the boundary conditions in the $x$-direction. We 
have a node at $x=R$ because there is a conducting plate here. At $x=L$ we also 
must have a node due to the assumed periodicity in $L$, since at $x=0$ we also
have a conducting plate. Therefore, when calculating $W_{L-R}$, we must take
Dirichlet boundary conditions in the $x$-direction, not the periodic boundary 
condition \cite{20}. However, in the $L\to\infty$ limit this difference does
not matter. Indeed, under the Dirichlet boundary conditions, the continuum 
limit corresponds to the replacement
$$\frac{1}{L-R}\sum\limits_{n=0}^\infty \to\int\limits_0^\infty \frac{dk_x}{\pi},$$
and we will have
\begin{equation}
W_{L-R}=\frac{\hbar c}{(2\pi)^2\,\pi}\,L^2(L-R)\int\limits_0^\infty dk_x
\int\limits_{-\infty}^\infty dk_y\int\limits_{-\infty}^\infty dk_z \sqrt{k_x^2+k_y^2+
k_z^2}= 
\frac{\hbar c}{(2\pi)^3}\,L^2(L-R)\int d\vec{k} \,\sqrt{k_x^2+k_y^2+k_z^2}.
\label{eq12}
\end{equation} 
Note that in Casimir's original paper \cite{13} (and late in some textbooks, 
see e.g. \cite{21}), the Dirichlet boundary conditions are used in all three 
dimensions (assuming the volume $L^3$ bounded by perfectly conducting walls). 
However, more natural in the spirit of field quantization procedure in quantum 
field theory is the assumption of periodic boundary conditions in the $y$- and 
$z$-directions \cite{20,22}. Of course, the final result does not depend on 
this choice.  

Only the difference $\Delta W=W_R+W_{L-R}-W_L$, which shows how the energy of
electromagnetic zero-point fluctuations changes due to the introduction of 
conducting plates, has a definite physical meaning \cite{13}. According to
(\ref{eq9}), (\ref{eq11}) and (\ref{eq12}), this difference is
\begin{equation}
\Delta W=\frac{\hbar c L^2}{(2\pi)^2}\iint dk_y\,dk_z\left \{
\sum\limits_{n=(0)1}^\infty\sqrt{k_y^2+k_z^2+\left(\frac{\pi n}{R}\right)^2}- 
\frac{R}{2\pi}\int\limits_{-\infty}^\infty dk_x\,\sqrt{k_x^2+k_y^2+k_z^2}\right \}.
\label{eq13}
\end{equation} 
Introducing dimensionless variables $x$ and $\nu$ through
\begin{equation}
k_y^2+k_z^2=k_\perp^2=\left(\frac{\pi}{R}\right)^2x,\;\;k_x=\frac{\pi}{R}\,\nu,
\;\;dk_y\,dk_z=2\pi k_\perp dk_\perp=\left(\frac{\pi}{R}\right)^2dx,
\label{eq14}
\end{equation}
we get
\begin{equation}
\Delta W=\frac{\pi^2 \hbar c L^2}{4R^3}\int\limits_0^\infty dx \left \{
\sum\limits_{n=(0)1}^\infty\sqrt{x+n^2}\;\lambda\left(\frac{\pi}{k_PR}\sqrt{x+n^2}
\right )- 
\frac{1}{2}\int\limits_{-\infty}^\infty d\nu\,\sqrt{x+\nu^2}\;
\lambda\left(\frac{\pi}{k_PR}\sqrt{x+\nu^2}\right )\right \},
\label{eq15}
\end{equation}
where, to take into account that real conductors are effectively transparent
above a certain frequency $\omega_P$ (plasma frequency of the metal) and, 
therefore, cannot provide ideal boundary conditions, we multiplied the 
contribution of each mode with the magnitude of the wave vector 
$k_n=\sqrt{k_y^2+k_z^2+\left(\frac{\pi n}{R}\right)^2}=\frac{\pi}{R}
\sqrt{x+n^2}$ by a cutoff function $\lambda(k/k_P)$, with the cut-off wave 
vector $k_P\sim\omega_P$, such that
\begin{equation}
\lambda(x)=\left\{\begin{array}{c}1,\;\;\mathrm{if}\;\;x\ll 1, \\
0,\;\;\mathrm{if}\;\;x\gg 1. \end{array}\right . 
\label{eq16}
\end{equation}
Thanks to this cutoff function, everything nicely converges in (\ref{eq15})
and we can calculate it using Euler-Maclaurin summation formula (see the 
appendix)
\begin{equation} 
\sum\limits_{n=0}^{\infty}f(n)-\int\limits_0^\infty f(x)\,dx=\frac{1}{2}\,f(0)-
\sum\limits_{n=2}^\infty \frac{B_n}{n!}\,\frac{d^{n-1}f(0)}{dx^{n-1}},
\label{eq17}
\end{equation}
where $B_n$ are Bernoulli numbers. To use this formula, we rewrite (\ref{eq15})
as follows
\begin{equation}
\Delta W=\frac{\pi^2 \hbar c L^2}{4R^3}\left\{\sum\limits_{n=(0)1}^\infty f(n)-
\frac{1}{2}\int\limits_{-\infty}^\infty d\nu\,f(\nu)\right \}=
\frac{\pi^2 \hbar c L^2}{4R^3}\left\{\ -\frac{1}{2}f(0)+\sum\limits_{n=0}^\infty 
f(n)-\int\limits_0^\infty d\nu\,f(\nu)\right \},
\label{eq18}
\end{equation}
where 
\begin{equation}
f(\nu)= \int\limits_0^\infty dx\sqrt{x+\nu^2}\,\lambda\left(\frac{\pi}{k_PR}\,
\sqrt{x+\nu^2}\right )= \int\limits_{\nu^2}^\infty dy\sqrt{y}\,\lambda\left
(\frac{\pi}{k_PR}\,\sqrt{y}\right ).
\label{eq19}
\end{equation}
To get (\ref{eq18}) from (\ref{eq15}), we have interchanged the order of 
summation and integrals, which is justified due to the absolute convergence 
in the presence of the cutoff function $\lambda$ \cite{23}.

Using the Leibniz rule for differentiating an integral (for generalizations 
of this useful formula to more than one dimension, see \cite{24})
\begin{equation}
\frac{d}{dt}\left (\;\int\limits_{g(t)}^{h(t)}F(x,t)\,dx\right )=F\left [h(t),t
\right ]\,\frac{dh}{dt}-F\left [g(t),t\right ]\,\frac{dg}{dt}+
\int\limits_{g(t)}^{h(t)}\frac{\partial F(x,t)}{\partial t}\,dx,
\label{eq20}
\end{equation}
we get
\begin{eqnarray} &&
\frac{df(0)}{d\nu}= \left. \left [ -2\nu^2\,\lambda\left(\frac{\pi}{k_PR}\,
\nu\right )\right ]\right \vert_{\nu=0}=0, \nonumber \\ && 
\frac{d^2f(0)}{d\nu^2}=\left . \left [-4\nu\lambda\left(\frac{\pi}{k_PR}\,
\nu\right )-2\nu^2\,\frac{\pi}{k_PR}\,\lambda^\prime\left(\frac{\pi}{k_PR}\,
\nu\right )\right ]\right \vert_{\nu=0}=0, \nonumber \\ && 
\frac{d^3f(0)}{d\nu^3}=\left [ -4\lambda\left(\frac{\pi}{k_PR}\,\nu\right )-
8\nu\,\frac{\pi}{k_PR}\,\lambda^\prime\left(\frac{\pi}{k_PR}\,\nu\right )-
\right . \nonumber \\ && \left . \left .
2\nu^2\,\left(\frac{\pi}{k_PR}\right)^2\,\lambda^{\prime\prime}
\left(\frac{\pi}{k_PR}\,\nu\right )\right ] \right \vert_{\nu=0}=-4.
\label{eq21}
\end{eqnarray}
It is clear that when $k>3$, $\frac{d^kf(0)}{d\nu^k}$ is proportional to
$\left(\frac{\pi}{k_PR}\right)^{k-3}$ and can be neglected as far as $k_PR
\gg 1$. Therefore, applying the Euler-Maclaurin summation formula in 
(\ref{eq18}), we recover the well-known result 
\begin{equation}
\Delta W=\frac{\pi^2\hbar c L^2}{R^3}\,\frac{B_4}{4!}=-\frac{\pi^2\hbar c L^2}
{720R^3}.
\label{eq22}
\end{equation} 
Since the pioneering work of Casimir \cite{13}, many other methods were 
proposed for solving the problem of zero-point fluctuations in the presence of 
boundaries that can alter the spectrum of zero-point oscillations (see, for 
example, \cite{15,18,25,26,27,28} and references therein).

Casimir force $F=-\frac{\partial \Delta W}{\partial R}$ that follows from
(\ref{eq22}) depends only on the fundamental constants $\hbar$ and $c$, and 
does not depend on  the  material  constitution of metallic plates. It is this 
universal nature of the Casimir effect that made it so famous.  

Remarkably, the Casimir force does not depend on the fine structure constant
$\alpha=e^2/(\hbar c)$. Naively, a dark photon, no matter how weakly it 
interacts with electrons, should double the Casimir force. However, 
the independence of the Casimir force from the fine structure constant is an 
illusion \cite{14}, and this can be seen from our derivation above.
 
The derivation of (\ref{eq22}) assumes $k_PR\gg 1$. But $k_P=\omega_P/c$, where
$\omega_P$ is the plasma frequency of the metal, and by introducing Bohr radius 
$a_B=\frac{\hbar c}{m_ec^2\alpha}$, $k_PR\gg 1$ condition can be rewritten in
the following way
\begin{equation}
\frac{\hbar \omega_P R}{\hbar c}=\frac{1}{\alpha}\,\frac{R}{a_B}\,\frac{\hbar
\omega_P}{m_e c^2}\gg 1.
\label{eq23}
\end{equation} 
For real metals, $\hbar\omega_P\approx 10~\mathrm{eV}$ \cite{29}, and 
experiments to measure the Casimir force are usually done at separations of 
the order of $R\approx 1~\mu\mathrm{m}$. For these numbers, the left-hand 
side of the inequality (\ref{eq23}) is about 50, and the inequality is 
well satisfied. However, if we decrease the electromagnetic coupling constant
$\alpha$ (but keep the electron mass $m_e$ intact), $\alpha a_B=\hbar/(mc)$
will not change. However, we will see in the next section that $\omega_P\sim
\sqrt{\alpha n}$, where $n\sim 1/a_B^3\sin\alpha^3$ is the number of 
conduction electrons per unit volume, and therefore $\omega_P\sim\alpha^2$. 
As a result, the left-hand side of the inequality (\ref{eq23}) decreases as 
$\alpha^2$, and at some point the inequality will be impossible to satisfy. 
In fact, it was argued in \cite{14} that in the limit $\alpha\to 0$ 
the Casimir force vanishes, like any other dynamical effect in quantum 
electrodynamics.  

It is sometimes assumed \cite{23} that all derivatives of the cutoff function 
$\lambda(x)$ vanish at the origin. Although such so-called flat functions 
do exist (for example, $\lambda(x)=1-e^{-1/x}$) and are frequently used in 
physics \cite{29A}, such a limitation of the permissible cutoff functions 
seems unnatural and devoid of physical motivation in the case of the Casimir 
effect. 
	
\section{Plasmons and dark photon}
To describe metals, we use the simple jellium model \cite{30}, according to 
which valence electrons are detached from atoms, and the result is a plasma, 
consisting of a collection of electrons and ions, which, on average, is 
electrically neutral. Such a model, of course, does not take into account the 
periodicity of the crystal lattice of a true solid, but in some respects this 
model is a good approximation to a real metal, which is a crystalline 
aggregate consisting of small crystals of random orientation. 

In addition, we use the self-consistent field approximation \cite{22,31,32}.
That is, we assume that electrons and ions  interact with electrostatic 
potentials $\phi(\vec{x})$ and $\tilde \phi(\vec{x})$, which are themselves 
determined by the Poisson equations, which depend on the  average charge 
densities in the plasma. Consequently, the dynamics of electrons and ions 
in plasma is determined by the (nonrelativistic) Hamiltonian \cite{22}
\begin{equation}
\hat H=\sum\limits_{s=e,i}\int d\vec{x}\psi^+_s\left[-\frac{\hbar^2}{2m_s}
\nabla^2+e_s\left(\sqrt{1-\epsilon^2}\phi-\epsilon\tilde\phi\right )\right ]
\psi_s=\hat H_0+\hat H_I,
\label{eq24}
\end{equation}
where the $s$ index was introduced to take into account various types of 
particles in the plasma (in our case, electrons and ions: $e_e=-e,e_i=e$). 
Expanding $\psi_s$ and  $\psi^+_s$ in free particle wave functions
\begin{equation}
\psi_s=\sum\limits_{\vec{q}} b_{s\vec{q}}\;\frac{e^{i\vec{q}\cdot\vec{x}}}
{\sqrt{\Omega}},\;\;\; \psi^+_s=\sum\limits_{\vec{q}} b^+_{s\vec{q}}\;
\frac{e^{-i\vec{q}\cdot\vec{x}}}{\sqrt{\Omega}},
\label{eq25}
\end{equation} 
where $\Omega$ is the quantization volume, we get
\begin{equation}
\begin{aligned}
&\hat H_0=\sum\limits_{s,\vec{q}}\frac{\hbar^2\vec{q}^{\,2}}{2m_s}\,
b^+_{s\vec{q}}\,b_{s\vec{q}},\;\;\;\hat H_I=\sum\limits_{s,\vec{q_1},\vec{q_2}} e_s\,
b^+_{s\vec{q}_1}\,b_{s\vec{q}_2}\,\phi^\prime(\vec{q}_1-\vec{q}_2,t),\\
&\phi^\prime(\vec{q},t)=\int\frac{d\vec{x}}{\Omega}\,e^{-i\vec{q}\cdot\vec{x}}\left(
\sqrt{1-\epsilon^2}\,\phi(\vec{x},t)-\epsilon\,\tilde\phi(\vec{x},t)\right ).
\end{aligned}
\label{eq26}
\end{equation}
The distribution function $F_s(\vec{q}^{\,\prime},\vec{q},t)$ is obtained by
quantum-mechanical and statistical averaging of the number operator
$b^+_{s\vec{q}^{\,\prime}}(t) b_{s\vec{q}}(t)$ \cite{22,32}:
\begin{equation}
F_s(\vec{q}^{\,\prime},\vec{q},t)=\sum\limits_\alpha P_\alpha \langle\alpha \vert
b^+_{s\vec{q}^{\,\prime}}(t) b_{s\vec{q}}(t)\vert\alpha \rangle,
\label{eq27}
\end{equation}
where $P_\alpha$ is the probability of finding the system in the quantum state
$\alpha$. 

Using standard anti-commutation relations
\begin{equation}
\{b_{s_1\vec{q}_1},\,b^+_{s_2\vec{q}_2}\}=\delta_{s_1s_2}\,\delta_{\vec{q}_1\vec{q}_2},\;\;\;
\{b_{s_1\vec{q}_1},\,b_{s_2\vec{q}_2}\}=0,\;\;\;
\{b^+_{s_1\vec{q}_1},\,b^+_{s_2\vec{q}_2}\}=0,
\label{eq28}
\end{equation}
and Heisenberg equations of motion
\begin{equation}
i\hbar\,\frac{\partial}{\partial t}\,b^+_{s\vec{q}^{\,\prime}}(t) b_{s\vec{q}}(t)=
-[\hat H,\,b^+_{s\vec{q}^{\,\prime}}(t) b_{s\vec{q}}(t)],
\label{eq29}
\end{equation}
we get 
\begin{equation}
\begin{aligned}
&\frac{\partial}{\partial t}F_s(\vec{q}^{\,\prime},\vec{q},t)=\frac{i}{\hbar}
\left(E_{sq^\prime}-E_{sq}\right )F_s(\vec{q}^{\,\prime},\vec{q},t)+ \\
&\frac{i}{\hbar}\,e_s\,\sum\limits_{\vec{p}}\left[\phi^{\prime}\left(\vec{p}-
\vec{q}^{\,\prime},t\right )F_s(\vec{p},\vec{q},t)- 
\phi^{\prime}\left(\vec{q}-\vec{p},t\right )F_s(\vec{q}^{\,\prime},\vec{p},t)
\right ],
\end{aligned}
\label{eq30}
\end{equation}
where $E_{sq}=\frac{\hbar^2\vec{q}^{\,2}}{2m_s}$.
This is a quantum-mechanical analogue of the Vlasov equation known in plasma 
physics \cite{32}.

In equilibrium, the distribution functions $F_s$ do not depend on time, and 
the potential $\phi^\prime$ vanishes (as does the charge density induced by 
the distribution functions $F_s$). Then (\ref{eq30}) is reduced to
$\left(E_{q^\prime}-E_q\right )F_s(\vec{q}^{\,\prime},\vec{q},t)$ with the solution 
$F_s(\vec{q}^{\,\prime},\vec{q},t)=F_{s0}(\vec{q})\,\delta_{\vec{q}\,\vec{q}^{\,\prime}}$. 
Small oscillations about this equilibrium are described by distribution 
functions
\cite{22,32}
\begin{equation}
F_s(\vec{q}^{\,\prime},\vec{q},t)=F_{s0}(\vec{q}\,)\,\delta_{\vec{q}\,\vec{q}^{\,\prime}}+
F_{s1}(\vec{q}^{\,\prime},\vec{q}\,)\,e^{-i\omega t},
\label{eq31}
\end{equation}
where $F_{s1}(\vec{q}^{\,\prime},\vec{q})$, like $\phi^\prime(\vec{q},t)=
\phi^\prime(\vec{q}\,)\,e^{-i\omega t}$, are assumed to be small quantities.
We linearize (\ref{eq30}) with respect to these  small quantities and find
the corresponding solution  \cite{32}
\begin{equation}
F_{s1}(\vec{q}^{\,\prime},\vec{q}\,)=-e_s\,\frac{F_{s0}(\vec{q}\,)-
F_{s0}(\vec{q}^{\,\prime}\,)}{\hbar\omega-\left (E_{sq}-E_{sq^\prime}\right )}\,
\phi^\prime\left (\vec{q}-\vec{q}^{\,\prime}\right ).
\label{eq32}
\end{equation}
For the average number densities, using decompositions (\ref{eq25}) we get
\begin{equation}
\begin{aligned}
&\bar{n}_s(\vec{x},t)=\sum\limits_\alpha P_\alpha \langle\alpha \vert
\psi^+_s(\vec{x},t)\, \psi_s(\vec{x},t)\vert\alpha \rangle= \\
&\sum\limits_{\vec{q},\,\vec{q}^{\,\prime}}F_s(\vec{q}^{\,\prime},\vec{q},t)\,
\frac{e^{i(\vec{q}-\vec{q}^{\,\prime})\cdot\vec{x}}}{\Omega}=
\sum\limits_{\vec{q},\,\vec{p}}F_s(\vec{p},\vec{p}+\vec{q},t)\,
\frac{e^{i\vec{q}\cdot\vec{x}}}{\Omega},
\end{aligned}
\label{eq33}
\end{equation}
where in the last step we just renamed the momentum variables for future 
convenience. 

In the self-consistent field approximation, these  average number densities 
should be used in the Poisson equations ($\mu=m_Dc/\hbar$, where $m_D$ is the 
dark photon mass):
\begin{equation}
\Delta\phi(\vec{x},t)=-\sum\limits_s e_s\sqrt{1-\epsilon^2}\,
\bar{n}(\vec{x},t),\;\;\;
(\Delta-\mu^2)\tilde \phi(\vec{x},t)=-\sum\limits_s e_s \epsilon\,
\bar{n}(\vec{x},t).
\label{eq34}
\end{equation}
The part of $\bar{n}(\vec{x},t)$ due to $F_{s0}$ corresponds to the equilibrium
with zero potentials $\phi$ and $\tilde\phi$. This leads to the condition
\begin{equation}
\frac{1}{\Omega}\sum\limits_{s,\vec{p}}e_s\,F_{s0}(\vec{p}\,)=0.
\label{eq35}
\end{equation} 
The part due to $F_{s1}$ corresponds to small oscillations near the 
equilibrium, and when substituted into (\ref{eq34}), along with decompositions
\begin{equation}
\phi(\vec{x},t)=\sum\limits_{\vec{q}}\phi(\vec{q}\,)\,e^{i\vec{q}\cdot\vec{x}
-i\omega t},\;\;\;\tilde\phi(\vec{x},t)=\sum\limits_{\vec{q}}\tilde
\phi(\vec{q}\,)\,e^{i\vec{q}\cdot\vec{x}-i\omega t},
\label{eq36}
\end{equation} 
leads to equations
\begin{flalign}
&-\sum\limits_{\vec{q}}\vec{q}^{\,2}\phi(\vec{q}\,)=\sum\limits_{s,\vec{q},\vec{p}}
\frac{e_s^2\sqrt{1-\epsilon^2}}{\Omega}\,\frac{F_{s0}(\vec{q}+\vec{p}\,)-
F_{s0}(\vec{p}\,)}{\hbar\omega-\left (E_{s,\vec{q}+\vec{p}}-E_{s,\vec{p}}\right)}\,
\phi^\prime (\vec{q}\,), \nonumber \\
& -\sum\limits_{\vec{q}}\left (\vec{q}^{\,2}+\mu^2\right )\tilde \phi(\vec{q}\,)=
-\sum\limits_{s,\vec{q},\vec{p}}\frac{e_s^2\epsilon}{\Omega}\,
\frac{F_{s0}(\vec{q}+\vec{p}\,)-F_{s0}(\vec{p}\,)}{\hbar\omega-\left 
(E_{s,\vec{q}+\vec{p}}-E_{s,\vec{p}}\right)}\,\phi^\prime (\vec{q}\,).
\label{eq37}
\end{flalign}
From this system of equations it is clear that 
\begin{equation}
\sum\limits_{\vec{q}}\left [\epsilon \,\vec{q}^{\,2}\phi(\vec{q}\,)+
\sqrt{1-\epsilon^2}\left (\vec{q}^{\,2}+\mu^2\right )\tilde \phi(\vec{q}\,)
\right ]=0.
\label{eq38}
\end{equation}
Therefore
\begin{equation}
\tilde \phi(\vec{q}\,)=-\frac{\epsilon}{\sqrt{1-\epsilon^2}}\,
\frac{\vec{q}^{\,2}}{\vec{q}^{\,2}+\mu^2}\,\phi(\vec{q}\,),
\label{eq39}
\end{equation}
and
\begin{equation}
\phi^\prime(\vec{q}\,)=\sqrt{1-\epsilon^2}\,\phi(\vec{q}\,)-\epsilon\,
\tilde \phi(\vec{q}\,)=\frac{1}{\sqrt{1-\epsilon^2}}\left[1-\frac{\epsilon^2
\mu^2}{\vec{q}^{\,2}+\mu^2}\right ]\phi(\vec{q}\,)= 
-\frac{\vec{q}^{\,2}+\mu^2}{\epsilon\,\vec{q}^{\,2}}\left[1-\frac{\epsilon^2
\mu^2}{\vec{q}^{\,2}+\mu^2}\right ]\tilde \phi(\vec{q}\,).
\label{eq40}
\end{equation}
Taking these relations into account, the system (\ref{eq37}) can be rewritten
as follows
\begin{equation}
\phi(\vec{q}\,)\,\epsilon(\vec{q},\omega)=0,\;\;\;
\tilde \phi(\vec{q}\,)\,\epsilon(\vec{q},\omega)=0,
\label{eq41}
\end{equation}
where $\epsilon(\vec{q},\omega)$, the plasma dielectric function \cite{22}, 
has the form
\begin{equation}
\epsilon(\vec{q},\omega)=1+\sum\limits_{s,\vec{p}}\frac{e_s^2}{\vec{q}^{\,2}
\Omega}\left[1-\frac{\epsilon^2\mu^2}{\vec{q}^{\,2}+\mu^2}\right ]
\frac{F_{s0}(\vec{q}+\vec{p}\,)-
F_{s0}(\vec{p}\,)}{\hbar\omega-\left (E_{s,\vec{q}+\vec{p}}-E_{s,\vec{p}}\right)}.
\label{eq42}
\end{equation}
At this point it is convenient \cite{22} to let the quantization volume 
$\Omega$ become infinite and replace $F_{s0}(\vec{p}\,)$ by the corresponding  
velocity distribution function $f_{s0}(\vec{V}\,)$, such that
\begin{equation}
\frac{1}{\Omega}\sum\limits_{\vec{p}}F_{s0}(\vec{p}\,)=\int\frac{d\vec{p}}
{(2\pi)^3}\,F_{s0}(\vec{p}\,)=\int f_{s0}(\vec{V}\,)\,d\vec{V}.
\label{eq43}
\end{equation}
Since $\vec{V}=\hbar\vec{p}/m$, we will have $d\vec{V}=\hbar^3d\vec{p}/m^3$
and therefore
\begin{equation}
f_{s0}(\vec{V}\,)=\frac{m^3}{(2\pi)^3\hbar^3}\,F_{s0}(\vec{q}\,),\;\;
\frac{1}{\Omega}\sum\limits_{\vec{p}} \to \int \frac{m^3}{(2\pi)^3\hbar^3}\,
d\vec{V}.
\label{eq44}
\end{equation}
Besides
\begin{equation}
E_{s,\vec{q}+\vec{p}}-E_{s,\vec{p}}=\frac{m_s}{2}\left [\left (
\vec{V}+\frac{\hbar \vec{q}}{m_s}\right )^2-\vec{V}^{\,2}\right ]=\frac{\hbar}
{2}\left [2\vec{V}\cdot\vec{q}+\frac{\hbar \vec{q}^{\,2}}{m_s}\right ],
\label{eq45}
\end{equation}
and (\ref{eq42}) takes the form
\begin{equation}
\epsilon(\vec{q},\omega)=1+\sum\limits_{s}\frac{e_s^2}{\hbar 
\vec{q}^{\,2}}\left[1-\frac{\epsilon^2\mu^2}{\vec{q}^{\,2}+\mu^2}\right ]
\int d\vec{V}\,\frac{f_{s0}\left(\vec{V}+\frac{\hbar\vec{q}}{m_s}\right )-
f_{s0}(\vec{V}\,)}{\omega-\vec{V}\cdot\vec{q}-\frac{\hbar\vec{q}^{\,2}}
{2m_s}}.
\label{eq46}
\end{equation}
In the classical limit $\hbar\to 0$, and
\begin{equation}
f_{s0}\left(\vec{V}+\frac{\hbar\vec{q}}{m_s}\right )-f_{s0}(\vec{V}\,)\approx
\sum\limits_{i=1}^3\frac{\hbar q_i}{m_s}\,\frac{\partial f_{s0}}{\partial V_i}
\equiv \frac{\hbar}{m_s}\,\vec{q}\cdot\nabla_Vf_{s0}.
\label{eq47}
\end{equation}
Therefore, in this limit,
\begin{equation}
\epsilon(\vec{q},\omega)=1+\sum\limits_{s}\frac{e_s^2}{m_s 
\vec{q}^{\,2}}\left[1-\frac{\epsilon^2\mu^2}{\vec{q}^{\,2}+\mu^2}\right ]
\int d\vec{V}\,\frac{\vec{q}\cdot\nabla_Vf_{s0}}{\omega-\vec{V}\cdot\vec{q}}.
\label{eq48}
\end{equation}
For the Maxwellian velocity distribution function
\begin{equation}
f_{s0}(\vec{V}\,)=\frac{n_s}{\pi^{3/2}\alpha_s^3}\,e^{-\vec{V}^{\,2}/\alpha_s^2},\;\;
\alpha_s=\sqrt{\frac{2kT}{m_s}},
\label{eq49}
\end{equation}
where $n_s=n$ is the density of $s$-type particles in the plasma assumed 
to be the same for electrons and ions, we have
\begin{equation}
\vec{q}\cdot\nabla_Vf_{s0}=-\frac{2n}{\pi^{3/2}\alpha_s^5}\,\vec{V}\cdot\vec{q}\,
e^{-\vec{V}^{\,2}/\alpha_s^2}.
\label{eq50}
\end{equation}
Assuming that $z$-axis is along $\vec{q}$, and using
\begin{equation}
\int\limits_{-\infty}^\infty e^{-x^2/\alpha^2}\,dx=\alpha\,\sqrt{\pi},\;\;
\frac{qV_z}{\omega-qV_z}=-1+\frac{\omega}{\omega-qV_z},
\label{eq51}
\end{equation}
we easily get
\begin{equation}
\epsilon(\vec{q},\omega)=1+\sum\limits_{s}\frac{e_s^2}{m_s 
q^2}\left[1-\frac{\epsilon^2\mu^2}{q^2+\mu^2}\right ]
\left [\frac{2n}{\alpha_s^2}-\frac{2n\omega}{\sqrt{\pi}\alpha_s^3q}\,
Z(z_s)\right ],
\label{eq52}
\end{equation}
where
\begin{equation}
q=\vert\vec{q}\vert,\;\;z_s=\frac{\omega}{q\alpha_s},\;\;
Z(z_s)=\int\limits_{-\infty}^\infty\frac{e^{-x^2}}{z_s-x}\,dx.
\label{eq53}
\end{equation}
Plasmon oscillations (Langmuir waves) correspond to the situation when
$z_e\gg 1$. Since $z_i/z_e=\alpha_e/\alpha_i=\sqrt{m_i/m_e}$, $z_e\gg 1$
necessarily implies $z_i\gg 1$, and we can use the following asymptotic
expansion of the Fried-Conte function $Z(z_s)$:
\begin{equation}
Z(z_s)\approx \frac{1}{z_s}\int\limits_{-\infty}^\infty e^{-x^2}\left (
1+\frac{x}{z_s}+\frac{x^2}{z_s^2}+\frac{x^3}{z_s^3}+\frac{x^4}{z_s^4}
\right )dx=\sqrt{\pi}\left (\frac{1}{z_s}+
\frac{1}{2z_s^3}+\frac{3}{4z_s^5}\right ).
\label{eq54}
\end{equation} 
As a result, we get
\begin{equation}
\epsilon(\vec{q},\omega)=1-\sum\limits_{s}\frac{e_s^2}{m_s 
q^2}\left[1-\frac{\epsilon^2\mu^2}{q^2+\mu^2}\right ]
\frac{n}{\alpha_s^2}\left (\frac{1}{z_s^2}+\frac{3}{2z_s^4}\right )= 
1-\sum\limits_{s}\frac{\omega_{ps}^2}{\omega^2}\left[1-
\frac{\epsilon^2\mu^2}{q^2+\mu^2}\right ]\left [1+\frac{3\alpha_s^2}
{2\omega^2}\,q^2\right ],
\label{eq55}
\end{equation} 
where \footnote{Remember that the  $e_s$ charge is expressed in 
Heaviside-Lorentz units. If $e_s$ is expressed in Gaussian units common in 
plasma physics, we must replace $e_s$ with $\sqrt{4\pi}e_s$ in the formulas.}
\begin{equation}
\omega_{ps}=\sqrt{\frac{n e_s^2}{m_s}}
\label{eq56}
\end{equation} 
are characteristic frequencies for electrons ($s=e$), and ions ($s=i$).
Using $\omega_{pi}^2=\omega_{pe}^2(m_e/m_i)$ and $\alpha_i^2=\alpha_e^2(m_e/m_i)$,
we finally get
\begin{equation}
\epsilon(\vec{q},\omega)=1-\left[1-\frac{\epsilon^2\mu^2}{q^2+\mu^2}\right ]
\left [\frac{\omega_{pe}^2}{\omega^2}\left(1+\frac{m_e}{m_i}\right )+
\frac{3\omega_{pe}^2\alpha_e^2\,q^2}{2\omega^4}\left(1+\frac{m_e^2}{m_i^2}\right )
\right ].
\label{eq57}
\end{equation}
A similar result is obtained if instead of the Maxwellian velocity 
distribution function (\ref{eq49}) we use the Fermi-Dirac distribution 
function
\begin{equation}
f_{s0}(\vec{V}\,)=\left \{ \begin{array}{c} \frac{3n}{4\pi V_{Fs}^3}, \;\;
\mathrm{if}\;\;V<V_{Fs} \\ 0,\quad\quad \;\;\;\mathrm{if}\;\;V>V_{Fs}.
\end{array}\right .
\label{eq58}
\end{equation}
In this case the thermal velocity $\alpha_e$ will be replaced by the Fermi
velocity $V_{Fe}$ and the final result will look like (for $\epsilon=0$ this
result was obtained in \cite{22,33})
\begin{equation}
\epsilon(\vec{q},\omega)=1-\left[1-\frac{\epsilon^2\mu^2}{q^2+\mu^2}\right ]
\left [\frac{\omega_{pe}^2}{\omega^2}\left(1+\frac{m_e}{m_i}\right )+
\frac{3\omega_{pe}^2V_{Fe}^2\,q^2}{5\omega^4}\left(1+\frac{m_e^3}{m_i^3}\right )
\right ].
\label{eq59}
\end{equation}
Equation (\ref{eq41}) shows that $\phi(\vec{q}\,)$ (as well as $\tilde 
\phi(\vec{q}\,)$) does not vanish only when $\epsilon(\vec{q},\omega)=0$.
In the long-wavelength approximation ($\alpha_e^2\,q^2\ll \omega_{pe}^2$),
an approximate solution of this equation has the form  
\begin{equation}
\omega^2\approx \left(1-\frac{\epsilon^2\mu^2}{q^2+\mu^2}\right )
\left(1+\frac{m_e}{m_i}\right )\omega_{pe}^2+\frac{3\left(1+\frac{m^2_e}{m^2_i}
\right )}{2\left(1+\frac{m_e}{m_i}\right )}\,\alpha_e^2\,q^2.
\label{eq60}
\end{equation}
As we see, for ultralight dark photon ($\mu^2\ll q^2$), modification of the
plasmon dispersion relation is negligible. In fact it is negligible for 
other values of $\mu^2$ also, since the natural magnitude of the mixing 
parameter $\epsilon$ is $\epsilon <10^{-2}-10^{-3}$ \cite{5}. The same 
conclusion is valid if we use Fermi-Dirac plasma dielectric function
(\ref{eq59}). In this case, an approximate solution of the $\epsilon(\vec{q},
\omega)=0$ equation has the form
\begin{equation}
\omega^2\approx \left(1-\frac{\epsilon^2\mu^2}{q^2+\mu^2}\right )
\left(1+\frac{m_e}{m_i}\right )\omega_{pe}^2+\frac{3\left(1+\frac{m^3_e}{m^3_i}
\right )}{5\left(1+\frac{m_e}{m_i}\right )}\,V_{Fe}^2\,q^2.
\label{eq61}
\end{equation}

The Fried-Conte function is defined in (\ref{eq53}) by a singular integral, and
its full specification requires a rule how to handle this singularity. Above
we assumed that the integral is given by the Cauchy principal value. Landau
argued \cite{34} that more correct way to deal with singularity in (\ref{eq53})
is to assume that the frequency $\omega$ (and hence $z_s$) has an infinitesimal
positive imaginary part: $z_s\to z_s+i\varepsilon$. This imaginary part leads 
to the so-called Landau damping \cite{35}\footnote{It seems,  A.A. Vlasov was
the first who anticipated the possibility of damping in a collisionless plasma 
\cite{36,37,38}.}. Using the Sokhotski-Plemelj formula
\begin{equation}
\lim_{\varepsilon\to 0}\frac{1}{x\pm i\varepsilon}=P\,\frac{1}{x}\mp i\pi\delta(x),
\label{eq62}
\end{equation}
we see that in our case the damping is proportional to $e^{-z^2_s}$ and, thus,
it  can be neglected, since $z_s\gg 1$.

\section{Dark light penetration depth in metals}
Euler-Lagrange equations that follow from the Lagrangian (\ref{eq4}) are
\begin{equation}
\partial_\nu F^{\mu\nu}=-e\sqrt{1-\epsilon^2}J^\mu,\;\;
\partial_\nu \tilde F^{\mu\nu}=\mu^2\tilde A^\mu+e\epsilon J^\mu-
\frac{e_b}{\sqrt{1-\epsilon^2}}\tilde J^\mu.
\label{eq63}
\end{equation}
We assume Lorenz conditions $\partial_\mu  A^\mu=\partial_\mu\tilde A^\mu=0$
for $A^\mu$ and $\tilde A^\mu$ four-potentials. For $A^\mu$ this is just a gauge 
fixing, while for $\tilde A^\mu$ it follows from the corresponding 
Euler-Lagrange equation in (\ref{eq63}), if the currents $J^\mu$ and 
$\tilde J^\mu$ are conserved and $\mu\ne 0$.

After the gauge fixing, the equations of motion (\ref{eq63}) for fields $A^\mu$
and $\tilde A^\mu$ become
\begin{equation}
\Box A^\mu=e\sqrt{1-\epsilon^2}J^\mu,\;\;\;
\left (\Box +\mu^2\right )\tilde A^\mu=-e\epsilon J^\mu+
\frac{e_b}{\sqrt{1-\epsilon^2}}\tilde J^\mu.
\label{eq64}
\end{equation}
Since electrons carry both the ordinary charge $-\sqrt{1-\epsilon^2}e$ and
the dark charge $\epsilon\,e$, Ohm's law inside a metal with conductivity 
$\sigma$ takes the form \cite{39}
\begin{equation}
eJ^\mu=\sigma\left (\sqrt{1-\epsilon^2}\,\vec{E}-\epsilon\,\vec{\tilde E}
\,\right ).
\label{eq65}
\end{equation}
Consider a transverse dark light plane wave propagating along the $z$ axis 
and entering a metal. We assume that the polarization of the electric field 
is along the $x$ axis, so that
\begin{equation}
A^\mu(\vec{x},t)=\left (0,A(z)\,e^{-i\omega t},0,0\right ),\;\;\;
\tilde A^\mu(\vec{x},t)=\left (0,\tilde A(z)\,e^{-i\omega t},0,0\right )
\label{eq66}
\end{equation}
Then
\begin{equation}
\begin{aligned}
&\vec{E}=-\frac{\partial \vec{A}}{\partial t}-\nabla\phi=
\left (i\omega A(z)\,e^{-i\omega t},0,0\right ),\\
&\vec{\tilde E}=-\frac{\partial \vec{\tilde A}}{\partial t}-\nabla\tilde\phi=
\left (i\omega \tilde A(z)\,e^{-i\omega t},0,0\right ),
\end{aligned}
\label{eq67}
\end{equation}
and the system (\ref{eq64}) takes the form (we have assumed that the dark 
current $\tilde J^\mu$ vanishes inside the metal)
\begin{flalign}
& \left (\partial_z^2+\omega^2\right )A(z)=-i\sigma\omega\sqrt{1-\epsilon^2}
\left [\sqrt{1-\epsilon^2}A(z)-\epsilon \tilde A(z) \right ],\nonumber \\
&  \left (\partial_z^2+\omega^2-\mu^2 \right )\tilde A(z)=i\sigma\omega
\epsilon
\left [\sqrt{1-\epsilon^2}A(z)-\epsilon \tilde A(z) \right ].
\label{eq68}
\end{flalign} 
The eigenmodes of this coupled system have the form $A(z)=C\,e^{ikz},\,
\tilde A(z)=\tilde C\,e^{ikz}$, where the eigenvalues $k$ are such 
that the homogeneous system 
\begin{eqnarray} &&
\left [-k^2+\omega^2+i\sigma\omega(1-\epsilon^2)\right ] C -
i\sigma\omega\epsilon\sqrt{1-\epsilon^2}\,\tilde C  = 0, \nonumber \\ &&
-i\sigma\omega\epsilon\sqrt{1-\epsilon^2}\,C +  \left [-k^2+\omega^2-\mu^2
+i\sigma\omega\epsilon^2\right ]\tilde C =0.
\label{eq69}
\end{eqnarray}
has non-zero solutions for $C,\,\tilde C$. Therefore, the determinant of the 
coefficient matrix of this system must be zero, and this condition leads to
the equation
\begin{equation}
(\omega^2-k^2)^2-\left [\mu^2-i\sigma\omega\right ](\omega^2-k^2)-
i\sigma\omega\mu^2(1-\epsilon^2)=0.
\label{eq70}
\end{equation}
Consequently, the eigenvalues are
\begin{eqnarray} &&
k_1^2=\omega^2-\frac{1}{2}\left (\mu^2-i\sigma\omega\right )+
\frac{1}{2}\sqrt{\left (\mu^2+i\sigma\omega\right )^2-4i\sigma\omega\epsilon^2
\mu^2}, \nonumber \\ &&
k_2^2=\omega^2-\frac{1}{2}\left (\mu^2-i\sigma\omega\right )-
\frac{1}{2}\sqrt{\left (\mu^2+i\sigma\omega\right )^2-4i\sigma\omega\epsilon^2
\mu^2}
\label{eq71}
\end{eqnarray}
Since $\epsilon$ is assumed to be small, we have approximately
\begin{equation} 
k_1^2\approx \omega^2+i\sigma\omega-\frac{i\sigma\omega\epsilon^2\mu^2}
{\mu^2+i\sigma\omega}= 
\omega^2 \left (1-\frac{\sigma^2\epsilon^2\mu^2}
{\mu^4+\sigma^2\omega^2}\right )+i\sigma\omega\left (1-\frac{\epsilon^2\mu^4}
{\mu^4+\sigma^2\omega^2}\right ), 
\label{eq72-1}
\end{equation}
and 
\begin{equation} 
k_2^2\approx \omega^2-\mu^2+\frac{i\sigma\omega\epsilon^2\mu^2}
{\mu^2+i\sigma\omega}= 
\omega^2 \left (1+\frac{\sigma^2\epsilon^2\mu^2}
{\mu^4+\sigma^2\omega^2}\right )-\mu^2+\frac{i\sigma\omega\epsilon^2\mu^4}
{\mu^4+\sigma^2\omega^2}.
\label{eq72-2}
\end{equation}
The corresponding eigenvectors are
\begin{equation}
\left (\begin{array}{c} A_1 \\ \tilde A_1\end{array}\right )=
C\left (\begin{array}{c} 1 \\ \frac{\sigma\omega\epsilon}{-\sigma\omega+
i\mu^2}\end{array}\right ),\;\;\;
\left (\begin{array}{c} A_2 \\ \tilde A_2\end{array}\right )=
\tilde C\left (\begin{array}{c} \frac{-\sigma\omega\epsilon}{-\sigma\omega+
i\mu^2}\\ 1 \end{array}\right ).
\label{eq73}
\end{equation}
Therefore, inside the metal $A(z)$ and $\tilde A(z)$ fields propagate 
according to
\begin{equation} 
A(z)=Ce^{ik_1z}-\tilde C\,\frac{\sigma\omega\epsilon}{-\sigma\omega+i\mu^2}\,
e^{ik_2z}\;\;\;\tilde A(z)=C\,\frac{\sigma\omega\epsilon}{-\sigma\omega+i\mu^2}\,
e^{ik_1z}+\tilde Ce^{ik_2z}.
\label{eq74}
\end{equation}
Boundary conditions at $z=0$ determine the coefficients $C$ and $\tilde C$.
In particular, if the boundary conditions are $A(0)=0,\,\tilde A(0)=
\tilde A_0$, we get
\begin{eqnarray} &&
A(z)=\tilde A_0\,\frac{\sigma\omega\epsilon \left(-\sigma\omega+i\mu^2\right )}
{\left(-\sigma\omega+i\mu^2\right )^2+\sigma^2\omega^2\epsilon^2}\left [e^{ik_1z}-
e^{ik_2z}\right ]\approx 
\tilde A_0\,\frac{\sigma\omega\epsilon}{-\sigma\omega
+i\mu^2}\left [e^{ik_1z}-e^{ik_2z}\right ], \\ && 
\tilde A(z)=\tilde A_0\,\frac{\left(-\sigma\omega+i\mu^2\right )^2}
{\left(-\sigma\omega+i\mu^2\right )^2+\sigma^2\omega^2\epsilon^2}
\left [e^{ik_2z}+\frac{\sigma^2\omega^2\epsilon^2}{\left (-\sigma\omega+
i\mu^2\right )^2}\,e^{ik_1z}\right]\approx \tilde A_0 e^{ik_2z},
\nonumber
\label{eq75}
\end{eqnarray}
and if the boundary conditions are $A(0)=A_0,\,\tilde A(0)=0$, then
\begin{eqnarray} &&
A(z)=A_0\,\frac{\left(-\sigma\omega+i\mu^2\right )^2}
{\left(-\sigma\omega+i\mu^2\right )^2+\sigma^2\omega^2\epsilon^2}
\left [e^{ik_1z}+\frac{\sigma^2\omega^2\epsilon^2}{\left (-\sigma\omega+
i\mu^2\right )^2}\,e^{ik_2z}\right] \approx A_0 e^{ik_1z}, \nonumber \\ && 
\tilde A(z)=A_0\,\frac{\sigma\omega\epsilon \left(-\sigma\omega+i\mu^2\right )}
{\left(-\sigma\omega+i\mu^2\right )^2+\sigma^2\omega^2\epsilon^2}\left [e^{ik_1z}-
e^{ik_2z}\right ]\approx  A_0\,\frac{\sigma\omega\epsilon}
{-\sigma\omega+i\mu^2}\left [e^{ik_1z}-e^{ik_2z}\right ].
\label{eq76}
\end{eqnarray}
As we can see, up to the leading terms in the $\epsilon$ expansion, if the 
incident plane wave is a pure dark light, ordinary vector-potential $A$ in
the metal will have the order of $\epsilon \tilde A$, and vice versa, if the 
incident plane wave is a pure ordinary light, the dark vector-potential 
$\tilde A$ in metal will be of the order of $\epsilon A$.

Equations (\ref{eq72-1}), (\ref{eq72-2}) show that $k_{1,2}^2$ and hence 
$k_{1,2}$ have positive imaginary parts. Then $e^{ikz}$ decreases exponentially 
in the metal. If the imaginary part is large, penetration depth (the so-called
skin depth \cite{40,41}) will be small. If $k^2=\rho e^{i\varphi}=
\rho\cos{\varphi}+i\rho\sin{\varphi}=a+ib$, with $\rho=\vert k^2\vert=
\sqrt{a^2+b^2}$, then $k=\sqrt{\rho} e^{i\varphi/2}=\sqrt{\rho}
\cos{(\varphi/2)}+i\sqrt{\rho}\sin{(\varphi/2)}=\alpha+i\beta$. 
In our case $b>0$ and, therefore,
$\varphi<\pi$. Hence $\cos{(\varphi/2)}=\sqrt{(1+cos{\varphi})/2}$,  
$\sin{\varphi/2}=\sqrt{(1-cos{\varphi})/2}$, and we get \cite{40}
\begin{equation}
\alpha=\sqrt{\frac{1}{2}\left (\sqrt{a^2+b^2}+a\right )},\;\;\;
\beta=\sqrt{\frac{1}{2}\left (\sqrt{a^2+b^2}-a\right )}.
\label{eq77}
\end{equation}
For the first mode $k_1^2=a_1+ib_1$ with $a_1\approx \omega^2$ and $b_1\approx 
\sigma\omega$. For good conductors $\sigma\gg\omega$ \cite{41}. Therefore
$b_1\gg a_1$ and from (\ref{eq77}) we get $\beta_1\approx\sqrt{b_1/2}$, which 
leads to the usual skin depth $\delta_1=1/\beta_1\approx \sqrt{\frac{2}
{\sigma\omega}}$ \cite{40}. 

In the case of the second mode $k_2^2=a_2+ib_2$, if ultralight dark photon 
$\mu^2\ll\omega^2$ is assumed, $a_2\approx \omega^2$ and $b_2\approx 
\frac{\epsilon^2\mu^4}{\sigma\omega}\ll a_2$. Therefore, $\beta_2\approx 
\frac{b_2}{2\sqrt{a_2}}=\frac{\epsilon^2\mu^4}{2\sigma\omega^2}$. The 
corresponding penetration depth is enormous:
\begin{equation}
\delta_2=\frac{\sigma\omega^2}{\epsilon^2\mu^4}=\frac{\delta_1}{\epsilon^2}
\left(\frac{\sigma}{\mu}\right)\left(\frac{\omega}{\mu}\right)^2
\sqrt{\frac{2\sigma\omega}{\mu^2}}\gg \delta_1.
\label{eq78}
\end{equation}
Consequently, the necessary boundary conditions cannot be realized, and  we 
conclude that the transverse polarizations of the ultralight dark photon do 
not make any significant contribution to the Casimir effect.

It remains to consider the propagation of the longitudinally polarized 
dark light. In this case we can take \cite{39}\footnote{In plasma, photons
acquire an effective mass \cite{42}, and therefore $A^\mu$ in a metal can have 
a nonzero longitudinal component.}
\begin{equation}
\begin{aligned}
&A^\mu\left(\vec{x},t\right )=\left(\phi(z)e^{-i\omega t},0,0,A(z)e^{-i\omega t}
\right ),\\
&\tilde A^\mu\left(\vec{x},t\right )=\left(\tilde \phi(z)e^{-i\omega t},0,0,
\tilde A(z)e^{-i\omega t}\right ).
\end{aligned}
\label{eq79}
\end{equation}
Taking into account the Lorenz conditions
\begin{equation}
-i\omega\phi(z)+\frac{dA(z)}{dz}=0,\;\;
-i\omega\tilde \phi(z)+\frac{d\tilde A(z)}{dz}=0,
\label{eq80}
\end{equation}
we get for electric fields
\begin{eqnarray} &&
E_z=-\frac{\partial}{\partial t}( A(z)e^{-i\omega t})-
\frac{\partial}{\partial z}( \phi(z) e^{-i\omega t}) =
-\frac{1}{i\omega}\left (\omega^2+\partial_z^2\right)A(z) e^{-i\omega t},
\nonumber \\ &&
\tilde E_z=-\frac{\partial}{\partial t}(\tilde A(z)e^{-i\omega t})-
\frac{\partial}{\partial z}(\tilde \phi(z) e^{-i\omega t})=
-\frac{1}{i\omega}\left (\omega^2+\partial_z^2\right)\tilde A(z)  e^{-i\omega t}.
\label{eq81}
\end{eqnarray}
Then from (\ref{eq64}), using  (\ref{eq65}) and assuming $A(z)=C\,e^{ikz},\,
\tilde A(z)=\tilde C\,e^{ikz}$, we get the following system
\begin{eqnarray} &&
\left (k^2-\omega^2\right )\left [1+i\frac{\sigma}{\omega}(1-\epsilon^2)
\right ] C-i\frac{\sigma}{\omega}\,\epsilon\sqrt{1-\epsilon^2}
\left (k^2-\omega^2\right )\tilde C=0,\nonumber \\ &&
-i\frac{\sigma}{\omega}\,\epsilon\sqrt{1-\epsilon^2}\left (k^2-\omega^2
\right )C +\left [\left (k^2-\omega^2\right )\left (1+i\frac{\sigma}{\omega}\,
\epsilon^2\right )+\mu^2\right ]\tilde C=0.
\label{eq82}
\end{eqnarray}
The corresponding eigenvalues  are
\begin{equation}
k_1=\omega,\;\;k_2^2=\omega^2-\mu^2\left (1-\frac{\epsilon^2\sigma^2}
{\sigma^2+\omega^2}\right )+i\sigma\omega\,\frac{\mu^2\epsilon^2}
{\sigma^2+\omega^2},
\label{eq83}
\end{equation}
with eigenvectors
\begin{equation}
\left (\begin{array}{c} A_1 \\ \tilde A_1\end{array}\right )=
C\left (\begin{array}{c} 1 \\ 0 \end{array}\right ),\;\;\;
\left (\begin{array}{c} A_2 \\ \tilde A_2\end{array}\right )=
\tilde C\left (\begin{array}{c} \frac{i\sigma\epsilon\sqrt{1-\epsilon^2}}
{\omega+i\sigma (1-\epsilon^2)}\\ 1 \end{array}\right ).
\label{eq84}
\end{equation}
Consequently, for $A(0)=0$, $\tilde A(0)=\tilde A_0$ boundary conditions, 
inside the metal 
\begin{equation}
A(z)=\tilde A_0  \frac{i\sigma\epsilon\sqrt{1-\epsilon^2}}
{\omega+i\sigma (1-\epsilon^2)}\left (e^{ik_2 z}-e^{ik_1 z}\right ),\;\;\;
\tilde A(z)=\tilde A_0 e^{ik_2 z}.
\label{eq85}
\end{equation}
In realistic case $\sigma\gg\omega\gg\mu$, $k_2^2\approx \omega^2+
i\frac{\omega}{\sigma}\mu^2\epsilon^2$ and (\ref{eq77}) will give for the
imaginary part of $k_2$ the result $\beta\approx \frac{\mu^2\epsilon^2}
{2\sigma}$. Therefore, the penetration depth for the longitudinal polarization
is 
\begin{equation}
\delta_L=\frac{2\sigma}{\mu^2\epsilon^2}=\frac{\delta_1}{\epsilon^2}
\left(\frac{\sigma}{\mu}\right)\sqrt{\frac{2\sigma\omega}{\mu^2}}\gg \delta_1.
\label{eq86}
\end{equation}
For an ultralight dark photon, $\delta_L\ll\delta_2$. However, the penetration 
depth of longitudinal polarization is still too large for the presence of
longitudinal polarization of a dark photon to have any appreciable effect on 
the Casimir force. 

\section{Leading contribution of ultralight dark photon to the Casimir energy}
It follows from the previous discussion that the leading contribution of 
ultralight dark photon to the Casimir energy arises from the modification of 
the plasma dielectric function $\sim \epsilon^2\mu^2$ and its effect on the 
propagation of $k_1^2\approx \omega^2+i\sigma\omega$ transverse mode.

One might think that the contribution of the longitudinal mode is also of the 
same order, since its penetration depth is of the order of $\sim \epsilon^2
\mu^2$. However, it was shown in \cite{45} that in the small-mass case the
contribution of longitudinal modes is actually of the order of $\sim
\mu^4$ and, thus, is insignificant in the present context. 

To calculate the contribution to the Casimir energy (per unit area) due to 
modification of the plasma dielectric function, we use the Lifshitz formula 
at zero temperature as an integral over the imaginary frequencies, in the form 
given in 
\cite{42A,42B} (remember, by default we use natural units $\hbar=1,\,c=1$):
\begin{equation}
w=\frac{1}{8\pi^2}\int\limits_0^\infty k dk \int\limits_{-\infty}^\infty d\xi
\left [\ln{\Delta_1(k,i\xi)}+\ln{\Delta_2(k,i\xi)}\right ],
\label{eq87}
\end{equation}
where
\begin{eqnarray} &&
\Delta_1(k,\omega)=1-\frac{(K_\epsilon-\epsilon(\omega)K)^2}{(K_\epsilon+
\epsilon(\omega)K)^2}\,e^{-2KR},\nonumber \\ &&
\Delta_2(k,\omega)=1-\frac{(K_\epsilon-K)^2}{(K_\epsilon+K)^2}\,e^{-2KR},
\label{eq88}
\end{eqnarray}
with
\begin{equation}
K^2=k^2-\omega^2,\;\;\;K_\epsilon^2=k^2-\epsilon(\omega)\,\omega^2.
\label{eq89}
\end{equation}
According to (\ref{eq57}), the plasma dielectric function at zero temperature
has the form
\begin{equation}
\epsilon(\vec{q},\omega)=1-\left[1-\frac{\epsilon^2\mu^2}{q^2+\mu^2}\right ]
\frac{\omega_{pe}^2}{\omega^2}.
\label{eq90}
\end{equation}
Taking $q^2=k_1^2\approx \omega^2+i\sigma\omega\approx i\sigma\omega$ in this
formula, we get
\begin{equation}
\epsilon(i\xi)=1+\left(\frac{\omega_{pe}}{\xi}\right)^2+\chi\left(
\frac{\omega_{pe}}{\xi}\right)^3\frac{1}{1-\frac{\mu^2}{\sigma\xi}},
\label{eq91}
\end{equation}
where
\begin{equation}
\chi=\frac{\epsilon^2\mu^2}{\sigma\omega_{pe}}\ll 1
\label{eq92}
\end{equation}
is a small parameter that determines the order of magnitude of the 
correction due to dark photons to the Casimir energy.

Consequently,
\begin{equation}
w(\chi)\approx w(0)+w^\prime(0)\,\chi,
\label{eq93}
\end{equation}
with
\begin{equation}
w(0)=\frac{1}{4\pi^2}\int\limits_0^\infty k dk \int\limits_0^\infty d\xi
\left [\ln{\Delta_1(k,i\xi)}+\ln{\Delta_2(k,i\xi)}\right ]_{\chi=0},
\label{eq87A}
\end{equation}
and
\begin{equation} 
w^\prime(0)=\frac{1}{8\pi^2}\int\limits_0^\infty k dk 
\int\limits_0^\infty d\xi \left(\frac{\omega_{pe}}{\xi}\right)^3 
\left(\frac{1}{1-\frac{\mu^2}{\sigma\xi}}+
\frac{1}{1+\frac{\mu^2}{\sigma\xi}}\right )  
\left [\frac{1}{\Delta_1}\,\frac{\partial \Delta_1}{\partial \epsilon}+
\frac{1}{\Delta_2}\,\frac{\partial \Delta_2}{\partial \epsilon}
\right ]_{\chi=0}, 
\label{eq94}
\end{equation}
where 
\begin{equation} 
\frac{\partial \Delta_1}{\partial \epsilon}=\frac{2K(K_\epsilon-\epsilon K)
(K_\epsilon^2+k^2)}{K_\epsilon(K_\epsilon+\epsilon K)^3}\,e^{-2KR},\;
\frac{\partial \Delta_2}{\partial \epsilon}=\frac{2\xi^4 K(1-\epsilon)}
{K_\epsilon(K_\epsilon+K)^4}\,e^{-2KR}.
\label{eq95}
\end{equation}
In (\ref{eq87}), considered as a double integral, we make the change of 
variables \cite{42A}
\begin{equation}
\xi=\frac{x}{2pR},\;\;\;k^2=\xi^2(p^2-1)=\frac{x^2}{4R^2}\left (1-\frac{1}{p^2}
\right ),
\label{eq96}
\end{equation}
with the Jacobian
\begin{equation}
\frac{\partial(k^2,\xi)}{\partial(p,x)}=\left \lvert \begin{array}{cc}
\frac{\partial k^2}{\partial p} & \frac{\partial k^2}{\partial x} \\
\frac{\partial \xi}{\partial p} & \frac{\partial \xi}{\partial x}
\end{array}\right \rvert=\frac{x^2}{4p^2R^3}.
\label{eq97}
\end{equation}
Then $K=p\xi$, $K_\epsilon=s\xi$, with $s=\sqrt{\epsilon-1+p^2}$, and we get
\begin{equation} 
w(0)=\frac{1}{32\pi^2R^3}\int\limits_0^\infty x^2 dx\int\limits_1^\infty \frac{dp}
{p^2} \left [\ln{\left (1-\frac{(s-p\epsilon)^2}{(s+p\epsilon)^2}\,
e^{-x}\right )}+\ln{\left (1-\frac{(s-p)^2}{(s+p)^2}\,e^{-x}\right )} 
\right ]_{\chi=0}.
\label{eq98}
\end{equation}
Note that
\begin{equation}
\epsilon\rvert_{\chi=0}=1+\frac{1}{\alpha^2},
\label{eq99}
\end{equation}
where
\begin{equation}
\alpha=\frac{\xi}{\omega_{pe}}=\frac{x}{2pR\omega_{pe}}=\frac{\delta_0}{R}\,
\frac{x}{2p},\;\;\delta_0=\frac{1}{\omega_{pe}},
\label{eq100}
\end{equation}
contains a small parameter $\delta_0/R\sim 0.1$ (for metals $\omega_{pe}\sim
10~\mathrm{eV}$, typical interplate separation in Casimir effect experiments 
is $R\sim 10^{-4}~\mathrm{cm}$ and we have used $\mathrm{eV}\cdot\mathrm{cm}
\approx 8066$).
In addition, due to the factor $e^{-x}$, the main contribution to the integral 
(\ref{eq98}) comes from the region of small $x$. Therefore, we can assume that
$\alpha$ is small in the integrand of (\ref{eq98}), and expand (up to first 
order in $\alpha$) 
\begin{equation} 
\begin{aligned}
&\left .\left (\frac{s-p}{s+p}\right )^2\right \rvert_{\chi=0}=
\left (\frac{\sqrt{1+\alpha^2p^2}
-\alpha p}{\sqrt{1+\alpha^2p^2}+\alpha p}\right )^2=\left (\sqrt{1+
\alpha^2p^2}-\alpha p\right )^4\approx 1-4\alpha p, \\
&\left .\left (\frac{s-p\epsilon}{s+p\epsilon}\right )\right \rvert_{\chi=0}^2=
\left (\frac{\alpha\sqrt{1+\alpha^2p^2}-(\alpha^2+1) p}
{\alpha\sqrt{1+\alpha^2p^2}+(\alpha^2+1) p}\right )^2\approx\left (
\frac{\alpha-p}{\alpha+p}\right )^2\approx1-\frac{4\alpha}{p}.
\end{aligned}
\label{eq101}
\end{equation}
Then (we substituted $\alpha=\frac{\delta_0}{R}\,\frac{x}{2p}$ in the 
final expressions below)
\begin{eqnarray} &&
\left .\ln{\left (1-\frac{(s-p\epsilon)^2}{(s+p\epsilon)^2}\,e^{-x}\right)}
\right \rvert_{\chi=0}
\approx -\ln{\frac{e^x}{e^x-1}}+\frac{2}{e^x-1}\,\frac{x}{p^2}\,
\frac{\delta_0}{R}, \nonumber \\ &&
\left .\ln{\left (1-\frac{(s-p)^2}{(s+p)^2}\,e^{-x}\right )}
\right \rvert_{\chi=0}\approx 
-\ln{\frac{e^x}{e^x-1}}+\frac{2}{e^x-1}\,\frac{\delta_0}{R}\,x,
\label{eq102}
\end{eqnarray}
and we get
\begin{equation}
\begin{aligned}
&w(0)=\frac{1}{32\pi^2R^3}\int\limits_0^\infty x^2 dx\int\limits_1^\infty 
\frac{dp}{p^2}\left[-2\ln{\frac{e^x}{e^x-1}}+\frac{2}{e^x-1}\,\frac{\delta_0}
{R}\,x\left(\frac{1}{p^2}+1\right )\right ]= \\
&\frac{1}{32\pi^2R^3}\int\limits_0^\infty x^2 dx\left[-2\ln{\frac{e^x}{e^x-1}}
+\frac{1}{e^x-1}\,\frac{\delta_0}{R}\,\frac{8x}{3}\right ].
\end{aligned}
\label{eq103}
\end{equation}
Performing integration by parts in the first term containing the logarithm,
and using the integral (see \cite{42C}, entry 3.411.1)
\begin{equation}
\int\limits_0^\infty \frac{x^3}{e^x-1}\,dx=\Gamma(4)\zeta(4)=\frac{\pi^4}{15},
\label{eq104}
\end{equation}
we finally get
\begin{equation}
w(0)=-\frac{1}{48\pi^2R^3}\left(1-4\,\frac{\delta_0}{R}\right )
\int\limits_0^\infty \frac{x^3}{e^x-1}\,dx=-\frac{\pi^2}{720R^3}
\left(1-4\,\frac{\delta_0}{R}\right ).
\label{eq105}
\end{equation}
The first term corresponds to the result (\ref{eq22}), and the second term is
the first order finite conductivity correction. It was calculated long ago 
(in 1961) by Dzyaloshinskiy, Lifshitz and Pitaevskii  with an error in the 
numerical coefficient corrected by Hargreaves in 1965 (see \cite{42B}). Here
we reproduce the correct coefficient just to gain a confidence in our 
calculational scheme. 

Finite conductivity corrections were calculated up to the fourth order in 
\cite{42A} and up to the sixth order in \cite{42D}. All of them are 
experimentally more important than the putative dark photon contribution, 
which we will now estimate. 

Similarly to what was done above, to calculate (\ref{eq94}) we expand the 
integrand in powers of $\alpha $, leaving only the leading term. 
Then equations (\ref{eq101}) indicate that
\begin{equation}
\Delta_1\rvert_{\chi=0}\approx \Delta_2\rvert_{\chi=0}\approx \frac{e^x-1}{e^x},
\label{eq106}
\end{equation}  
while
\begin{eqnarray} &&
\left . \frac{\partial \Delta_1}{\partial \epsilon}\right \rvert_{\chi=0}=
\frac{2p(s-\epsilon p)(s^2+p^2-1)}{s(s+\epsilon p)^3}\,e^{-x}\approx 
-\frac{2\alpha^3}{p}\,e^{-x}, \nonumber \\ &&
\left . \frac{\partial \Delta_2}{\partial \epsilon}\right \rvert_{\chi=0}=
\frac{2p(1-\epsilon )}{s(s+p)^4}\,e^{-x}\approx 
-2\alpha^3 p\,e^{-x}.
\label{eq107}
\end{eqnarray}
Therefore,
\begin{equation}
w^\prime(0)\approx -\frac{1}{32\pi^2R^3}\int\limits_0^\infty \frac{x^2}{e^x-1}\,dx
\int\limits_1^\infty \left (\frac{1}{p^3}+\frac{1}{p}\right )
\left (\frac{1}{1-\varkappa p/x}+\frac{1}{1+\varkappa p/x}\right ) dp,
\label{eq108}
\end{equation}
where
\begin{equation}
\varkappa=\frac{2\mu^2R}{\sigma}\ll 1,\;\;\;
\chi=\epsilon^2\,\frac{\delta_0}{2R}\,\varkappa.
\label{eq109}
\end{equation}
Integrals over $p$ can be done using the partial fraction decompositions:
\begin{eqnarray} &&
\frac{1}{p(1+\varkappa p/x)}=\frac{1}{p}-\frac{\varkappa}{x}\,\frac{1}
{(1+\varkappa p/x)}, \nonumber \\ &&
\frac{1}{p^3(1+\varkappa p/x)}=\frac{1}{p^3}-
\frac{\varkappa}{x}\,\frac{1}{p^2}+\frac{\varkappa^2}{x^2}\,\frac{1}{p}-
\frac{\varkappa^3}{x^3}\,\frac{1}{(1+\varkappa p/x)}.
\label{eq110}
\end{eqnarray}
As a result, we get
\begin{eqnarray} &&
w^\prime(0)\approx -\frac{1}{32\pi^2R^3}\int\limits_0^\infty\frac{x^2\,dx}
{e^x-1}\left [1-2\ln{\frac{\varkappa}{x}}+\left(1+\frac{\varkappa^2}{x^2}
\right )\ln{\left \lvert 1-\frac{\varkappa^2}{x^2}\right\rvert}\,\right ]
\approx \nonumber \\ &&
-\frac{1}{32\pi^2R^3}\int\limits_0^\infty\frac{x^2\left(1-2\ln{\varkappa}+
2\ln{x}\right)}{e^x-1}\,dx.
\label{eq111}
\end{eqnarray}
But (the evaluation of these integrals is explained in detail in appendix
\ref{App2})
\begin{equation}
\int\limits_0^\infty\frac{x^2}{e^x-1}\,dx=2\zeta(3),\;\;\;
\int\limits_0^\infty\frac{x^2\,\ln{x}}{e^x-1}\,dx=2\zeta^\prime(3)+(3-2\gamma)
\,\zeta(3),
\label{eq112}
\end{equation}
where $\gamma\approx 0.5772$ is the Euler constant, $\zeta(3)\approx 1.2021$
is the Ap\'{e}ry constant, and $\zeta^\prime(3)\approx -0.1981$. Consequently, 
we get
\begin{equation}
w^\prime(0)\approx -\frac{\zeta(3)}{16\pi^2R^3}\left (4-2\gamma+
\frac{\zeta^\prime(3)}{\zeta(3)}-2\ln{\varkappa}\right )\approx 
-\frac{\zeta(3)}{8\pi^2R^3}\left (1.34-\ln{\varkappa}\right )\approx
\frac{\zeta(3)}{8\pi^2R^3}\,\ln{\varkappa}.
\label{eq113}
\end{equation}
Therefore, our final result for the leading contribution of the ultralight 
dark photon to the Casimir energy is (we have restored $\hbar$ and $c$, and
scaled the energy in accordance with the area of the plates, as in 
(\ref{eq22}))
\begin{equation}
W_D\approx \frac{\zeta(3)\hbar c L^2}{8\pi^2R^3}\,\frac{\epsilon^2\mu^2}
{\sigma\omega_{pe}}\,\ln{\left(\frac{2\mu^2R}{\sigma}\right)}.
\label{eq114}
\end{equation}
In Drude's free electron theory of metals $\sigma=\frac{ne^2\tau}{m}$, 
where $\tau\approx 2\times 10^{-14}~\mathrm{sec}\approx\frac{1}
{0.2~\mathrm{eV}}$ is the average time between two consecutive electron 
collisions \cite{42E}. Comparing with (\ref{eq56}), we obtain (for 
$\omega_{pe}\sim 10~\mathrm{eV}$)
\begin{equation}
\sigma =\omega_{pe}^2\tau\approx 50\,\omega_{pe}\approx 500~\mathrm{eV}.
\label{eq115}
\end{equation}
Then, for $R\sim 10^{-4}~\mathrm{cm}\approx \frac{1}{1.24~\mathrm{eV}}$,
\begin{equation}
\sqrt{\frac{\sigma}{2R}}\approx 18~\mathrm{eV},
\label{eq116}
\end{equation}
and the correction due to dark photon to the Casimir energy can be 
represented in the form (the much more important finite conductivity 
corrections and all other non-dark photon corrections are omitted 
for clarity)
\begin{equation}
W\approx -\frac{\pi^2\hbar c L^2}{720 R^3}\left [1-\epsilon^2\,
\frac{90\,\zeta(3)}{\pi^4}\,\frac{\delta_0}{2R}\,\frac{2\mu^2 R}{\sigma}\,
\ln{\left (\frac{2\mu^2 R}{\sigma}\right )}\right]\approx 
-\frac{\pi^2\hbar c L^2}{720 R^3}\left [1-\epsilon^2\,\frac{90\,\zeta(3)}
{\pi^4}\,\frac{\delta_0}{R}\left(\frac{\mu}{18~\mathrm{eV}}\right )^2
\ln{\left(\frac{\mu}{18~\mathrm{eV}}\right )}\right ].
\label{eq117}
\end{equation}
It is clear from(\ref{eq109}) that our approximations are valid as far as 
$\mu\ll 18~\mathrm{eV}$.

The function $\varkappa^2\ln{\varkappa^2}$ reaches its minimum $-e^{-1}$ at
$\varkappa^2=e^{-1}$ (which corresponds to $\mu\sim 10~\mathrm{eV}$ for
$\sigma\sim  500~\mathrm{eV}$ and $R\sim 10^{-4}~\mathrm{cm}$). Therefore, it
follows from (\ref{eq117}) that the magnitude of the dark photon correction is
less than $2\times 10^{-2}\,\epsilon^2$. If dark photons comprise all of the
dark matter, hidden-photon search experiments indicate $\epsilon<10^{-12}$ in
the $\sim\mathrm{eV}$ mass range \cite{1EXP,2EXP}. Although it is possible 
that dark photons make up only a small fraction of the dark matter 
\cite{Alizzi}, the requirement that the Sun does not emit more than 10\% of 
its photon luminosity in a dark channel still drastically constraints the 
kinetic mixing parameter: $\epsilon<4\times 10^{-12}\,\frac{\mathrm{eV}}{\mu}$ 
\cite{sun1,sun2}. Therefore, we conclude that the dark photon contribution to 
the Casimir energy is negligible from the experimental point of view.   

\section{Conclusions}
The Casimir effect for two mixed scalar fields was analyzed in \cite{43}. It 
was shown that if the zero-point energy is evaluated for the vacuum of fields 
with definite mass, then the result is independent of the mixing parameters 
(in this case, the Casimir force is simply the sum of the Casimir forces for 
two unmixed fields). In contrast, if  the Casimir force is evaluated
using the flavor vacuum, the result shows an explicit dependence on the mixing 
parameters. Thus, these two approaches give different, albeit numerically 
close, results for the Casimir force, thus demonstrating a nonequivalence 
between mass and flavor representations for mixed fields \cite{43}.

However, in \cite{43} ideal boundary conditions were assumed. Our results
in this work show that the expected penetration depths for dark photons into
real metals are very large, so great that the necessary boundary conditions for 
the Casimir effect cannot be met. As a result, the presence of dark photons
cannot significantly affect the Casimir force \footnote{It has been 
experimentally demonstrated that the Casimir force between metallic films 
decreases significantly when the layer thickness is less than the skin depth, 
which for most common metals is about $10^{-8}~\mathrm{m}$ \cite{43A}}. 

At about $R\sim 10^{-4}~\mathrm{cm}$ interplate distances, the leading 
contribution to the Casimir energy arises from the modification of the 
frequency-dependent dielectric function of metals. We have estimated this
contribution, which for expected dark photon parameters turned out to be 
negligible.

An indirect way of how dark photons could affect the Casimir force could be 
a change in the properties of plasmons associated with the presence of dark 
photons, since it is known that at short distances the interaction between 
plasmon excitations at the surface of Casimir mirrors dominates in the real 
Casimir force \cite{44}. However, our results show that the expected 
modification of the plasmon dispersion relation is insignificant and, 
therefore, such an indirect effect of dark photons on the Casimir force is 
also negligible.
 	
Theoretically, the problem of the influence of field mixing on the Casimir 
force is very interesting, especially in light of the unitary
nonequivalence of the mass and flavor vacua. It's a pity that the expected 
effects are negligible and, therefore, the Casimir effect does not allow an 
experimental study of this interesting phenomenon.

In the course of this work, we assumed an ultralight dark photon. Note, 
however, that the main result, namely the insensitivity of the Casimir force 
to the presence of a dark photon for all practical purposes, remains valid 
for a massive dark photon. 

The Casimir effect for massive photons was investigated by Barton and Dombey
many years ago \cite{45}. As it is well-known \cite{46}, the Casimir force is
exponentially suppressed by the factor $e^{-2\mu R}$, where $\mu$ is a photon
mass and $R$ is the distance between the plates, for $\mu R\gg 1$. The 
frequencies that dominate the Casimir force are of the order of $1/R$ 
\cite{14,27,28}. Therefore, the condition $\mu R\sim 1$ is translated into 
the condition $\mu\sim \omega$ and for small mixing parameter $\epsilon$ 
(\ref{eq78}) and (\ref{eq86}) show that the penetration depths remain
unacceptably large. 
 
Finally, note that there are physics scenarios beyond the Standard 
Model that are different from the dark photon model. The implications of the 
Casimir effect for models such as universal extra dimensions, Randall-Sundrum 
models, and scale-invariant models have been discussed in \cite{47}.

\section*{Acknowledgments}
We would like to thank Carlo Beenakker and the {\it MathOverflow} user 
with the nickname dan\_fulea for suggesting the ideas that were used in the 
Appendix \ref{App2}. We also thank an anonymous reviewer for constructive 
comments. The work is supported by the Ministry of Education and Science of 
the Russian Federation.

\appendix
\section{Euler-Maclaurin Summation Formula}
\label{App1}
The Euler-Maclaurin summation formula was originally obtained in 1782 by Euler
and independently and almost simultaneously by Maclaurin \cite{1AA}. 
A rigorous treatment of this very important tool of numerical analysis with
diverse applications can be found, for example, in \cite{1A}. There exist
several elementary derivations \cite{1B,1C,1D} of this remarkable formula.
However, we prefer a formal heuristic derivation \cite{1BB}, which in the 
simplest way demystifies the appearance of Bernoulli numbers in it. Note that 
using Banach spaces of entire functions of exponential type, the formal 
derivation of the Euler-Maclaurin formula can be made mathematically 
rigorous \cite{1E}.

Let $\hat D=\frac{d}{dx}$ be a differential operator, so that
\begin{equation} 
\hat D f(x)=\frac{d f}{dx},\;\;\;f(x+n)=e^{n\hat D}f(x),
\label{eqA1}
\end{equation}
where the second equation is a formal expression of the Taylor formula.
Then
\begin{equation} 
\sum\limits_{n=0}^{N-1}f(x+n)=\left (\sum\limits_{n=0}^{N-1} e^{n\hat D}\right )f(x)=
\frac{e^{N\hat D}-1}{e^{n\hat D}-1}f(x)=\frac{1}{e^{n\hat D}-1}\left[f(x+N)-f(x)
\right ].
\label{eqA2}
\end{equation}
The Bernoulli numbers $B_n$ are defined by the power series expansion of their
exponential generating  function:
\begin{equation} 
\frac{x}{e^x-1}=\sum\limits_{n=0}^\infty \frac{B_n}{n!}\,x^n,\;\;\;
B_0=1,\;\;B_1=-\frac{1}{2},\;\;B_2=\frac{1}{6}, 
B_3=0,\;\;B_4=-\frac{1}{30},\;\;B_5=0,\;\;B_6=\frac{1}{42} \ldots
\label{eqA3}
\end{equation}
Comparing (\ref{eqA3}) and (\ref{eqA2}), we can write
\begin{equation} 
\sum\limits_{n=0}^{N-1}f(x+n)=\left (\hat D^{-1}+\sum\limits_{n=1}^\infty 
\frac{B_n}{n!}\,\hat D^{n-1}\right )\left [f(x+N)-f(x)\right ].
\label{eqA4}
\end{equation}
On the other hand,
\begin{equation} 
\int\limits_0^N f(x+t)\,dt=\int\limits_0^N e^{t\hat D}f(x)\,dt=
\hat D^{-1}\left (e^{N\hat D}-1\right )f(x)=\hat D^{-1}\left [f(x+N)-
f(x)\right ].
\label{eqA5}
\end{equation}
Therefore
\begin{equation} 
\sum\limits_{n=0}^{N-1}f(x+n)-\int\limits_0^N f(x+t)\,dt=\sum\limits_{n=1}^\infty 
\frac{B_n}{n!}\,\hat D^{n-1}\left [f(x+N)-f(x)\right ].
\label{eqA6}
\end{equation}
When $N\to\infty$, for convergence we need $\hat D^kf(\infty)=0$, $k=0,1,
\ldots$ and (\ref{eqA6}) in this limit takes the form
 \begin{equation} 
\sum\limits_{n=0}^{\infty}f(x+n)-\int\limits_0^\infty f(x+t)\,dt=
-\sum\limits_{n=1}^\infty \frac{B_n}{n!}\,\hat D^{n-1} f(x)=\frac{1}{2}\,f(x)-
\sum\limits_{n=2}^\infty \frac{B_n}{n!}\,\frac{d^{n-1}f(x)}{dx^{n-1}}\,.
\label{eqA7}
\end{equation}
The form of the Euler-Maclaurin Summation Formula used in the main text
corresponds to $x=0$ in (\ref{eqA7}).

\section{Evaluation of the integrals}
\label{App2}
\setcounter{equation}{0}
First we evaluate the integral from entry 3.411.1 in \cite{42C}.
From the definition of the gamma function
\begin{equation}
\Gamma(\nu)=\int\limits_0^\infty x^{\nu-1}e^{-x}dx,
\label{eqB1}
\end{equation}
and using
\begin{equation}
\frac{1}{1-e^{-x}}=\sum\limits_{k=0}^\infty e^{-kx},
\label{eqB2}
\end{equation}
we get after interchanging the orders of the integration and summation
\begin{eqnarray} &&
\int\limits_0^\infty \frac{x^{\nu-1}}{e^x-1}dx=\int\limits_0^\infty 
\frac{x^{\nu-1}e^{-x}}{1-e^{-x}}dx=\int\limits_0^\infty\sum\limits_{k=0}^\infty
x^{\nu-1}e^{-(k+1)x}dx= \nonumber \\ &&
\sum\limits_{k=0}^\infty\int\limits_0^\infty 
x^{\nu-1}e^{-(k+1)x}dx=\Gamma(\nu)\sum\limits_{k=0}^\infty \frac{1}
{(k+1)^n}=\Gamma(\nu)\,\zeta(\nu).
\label{eqB3}
\end{eqnarray}
Now, if we differentiate the just proved identity
\begin{equation}
\zeta(\nu)=\frac{1}{\Gamma(\nu)}\int\limits_0^\infty \frac{x^{\nu-1}}{e^x-1}
dx
\label{eqB4}
\end{equation}
with respect to $\nu$ and take into account $\frac{dx^{\nu-1}}{d\nu}=\ln{x}\,
x^{\nu-1}$, we get
\begin{equation} 
\zeta^\prime(\nu)=\frac{1}{\Gamma(\nu)}\int\limits_0^\infty\frac{x^{\nu-1}\,
\ln{x}}{e^x-1}dx-\frac{\Gamma^\prime(\nu)}{\Gamma^2(\nu)}
\int\limits_0^\infty\frac{x^{\nu-1}}{e^x-1}dx= 
\frac{1}{\Gamma(\nu)}\int\limits_0^\infty\frac{x^{\nu-1}\,
\ln{x}}{e^x-1}dx-\psi(\nu)\,\zeta(\nu),
\label{eqB5}
\end{equation}
where $\psi(\nu)=\Gamma^\prime(\nu)/\Gamma(\nu)$. Therefore,
\begin{equation}
\int\limits_0^\infty\frac{x^{\nu-1}\,\ln{x}}{e^x-1}dx=
\Gamma(\nu)\,\zeta^\prime(\nu)+\psi(\nu)\,\Gamma(\nu)\,\zeta(\nu).
\label{eqB6}
\end{equation}
Note that from $\psi(\nu+1)=\frac{1}{\nu}+\psi(\nu)$ and $\psi(1)=-\gamma$,
it follows that
\begin{equation}
\psi(3)=\frac{1}{2}+1+\psi(1)=\frac{1}{2}\left (3-2\gamma\right).
\label{eqB7}
\end{equation}

\end{document}